\documentclass[12pt,a4paper,dvips]{article}
\usepackage{a4p}
\usepackage{rotating}
\usepackage{cite,mcite}
\usepackage{graphicx}
\usepackage{epsfig}
\usepackage{physics}
\usepackage{l3_titlePH,ifthen}

\journalname{Phys. Lett. B}

\date{February 23, 2005}

\preprint{2005-006}

\newlength{\capindent}
\newlength{\capwidth}
\newlength{\figwidth}
\newcommand{\icaption}[2][!*!,!]{\hspace*{\capindent}%
  \begin{minipage}{\capwidth}
    \ifthenelse{\equal{#1}{!*!,!}}%
      {\caption{#2}}%
      {\caption[#1]{#2}}
  \end{minipage}}

\newcommand{\qqgg}{\ensuremath{\rm q\bar{q}\gamma\gamma}}

\begin{document}
%%%%%%%%%%%%%%%%%%%%%%%%%%%%%%%%%%%%%%%%%%%%%%%%%%%%%%%%%%%%%%%%%%%%%%%%%%%%%%%
% 

\begin{titlepage}

\title{Z-boson production with\\ two unobserved, back-to-back, hard
  photons at LEP}  

\author{L3 Collaboration}

\begin{abstract}

The double-radiative process $\epem\ra\Zo\gamma\gamma\rightarrow\qqgg $ where the two
hard photons escape  detection at low polar angles into opposite
directions, is studied in 0.62\,fb$^{-1}$
of data collected with the L3 detector at LEP at centre-of-mass
energies between $188.6\GeV$ and $209.2\GeV$.  The cross sections are
measured and found to be consistent with the Standard Model
expectations.

\end{abstract}

\submitted
\end{titlepage}

%
%%%%%%%%%%%%%%%%%%%%%%%%%%%%%%%%%%%%%%%%%%%%%%%%%%%%%%%%%%%%%%%%%%%%%%%%%%%%%%%
\section{Introduction}
%%%%%%%%%%%%%%%%%%%%%%%%%%%%%%%%%%%%%%%%%%%%%%%%%%%%%%%%%%%%%%%%%%%%%%%%%%%%%%%
%

One of the most copious sources of events in $\epem$ collisions at LEP
above the Z resonance is the process $\epem\ra\qqbar$, with a cross
section of about 100~pb. The effective centre-of-mass energy,
$\sqrt{s'}$, at which this hadron production takes place does not
necessarily correspond to the centre-of-mass energy of the LEP
machine, $\sqrt{s}$, owing to the emission of one or more
hard initial-state-radiation (ISR) photons by the incoming electrons or
positrons.  These photons are most likely emitted along the beam line,
in the low polar-angle regions of the detectors which are not
instrumented and, therefore, escape detection. 

The cross section of the $\epem\ra\qqbar$ process was
measured~\cite{l3fpp,othersFpp} and found to be in agreement with the
Standard Model predictions for both a subsample of events with values
of $\sqrt{s'}$ close to $\sqrt{s}$ and a more inclusive sample
extending to lower values of $\sqrt{s'}$.  The emission of ISR photons
often implies $\sqrt{s'}\approx m_{\rm Z}$, where $m_{\rm
Z}=91.19\GeV$ is the mass of the Z boson. This phenomenon is commonly
called ``radiative return to the Z\,''. The process
$\epem\ra\Zo\gamma\ra\qqbar\gamma$, where a hard ISR photon is
responsible for such radiative return to the Z, was studied in
detail\cite{L3nTGC,mZ}. Events in which the photon was visible in the
detector, were used to constrain possible anomalous triple-couplings
between neutral gauge bosons~\cite{L3nTGC}. Events with either a
detected photon or a low-angle undetected photon were used to
reconstruct the mass of the Z boson and validate the analysis tools
used in the measurement of the mass of the W boson\cite{mZ}. The
$\epem\ra\Zo\gamma\gamma\ra\qqbar\gamma\gamma$ process, where both ISR
photons were visible in the detector, was first observed by the L3
collaboration~\cite{L3ZggFirst}. The cross section of this process was
then measured for $\sqrt{s}=130-209\GeV$ and found to be in agreement
with the predictions~\cite{L3Zgg}.

This Letter extends the study of the
$\epem\ra\Zo\gamma\gamma\ra\qqbar\gamma\gamma$ process to the case in
which both ISR photons are emitted at low polar angles and are
therefore not detected. In particular, the case is considered in which
the two photons are emitted on opposite sides of the detector, with
comparable transverse momenta. In this topology, the two jets
originating from the Z-boson decay are back-to-back. In the following,
this process is denoted as ``double-radiative return to the Z\,''.
This study complements the previous studies of the $\epem\ra\qqbar$
process in a very specific phase-space region and allows further tests
of Monte Carlo simulations of ISR photons in hadronic
events. Moreover, final states with two back-to-back hadronic jets and
missing energy are a signature of the near-threshold production of the
Standard Model Higgs boson, H, at LEP in the reaction $\rm\epem\ra
ZH$. In this case, the missing energy is due to a Z boson decaying
into neutrinos and the jets to the Higgs boson. In addition,
manifestations of New Physics in the production of an
invisibly-decaying Higgs boson in association with a Z boson decaying
into hadrons would also give rise to the same final state. Finally,
similar event topologies are predicted by Supersymmetry. Therefore, a
study of double-radiative return to the Z with unobserved photons
validates the background Monte Carlo simulations for those searches.

\begin{figure}[p]
  \begin{center}
     \begin{tabular}{cc}
       \mbox{\includegraphics[width=.5\textwidth]{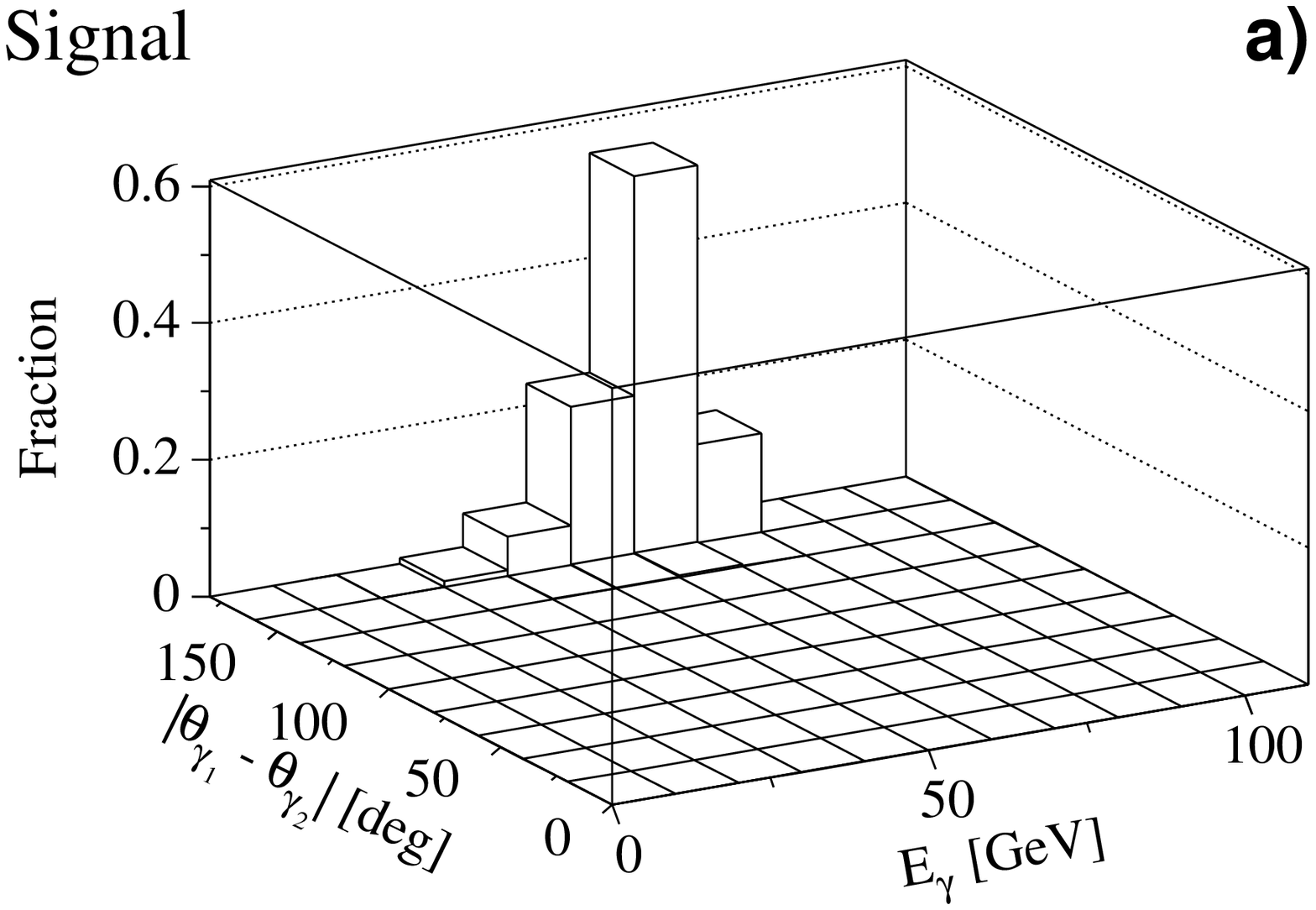}} &
       \mbox{\includegraphics[width=.5\textwidth]{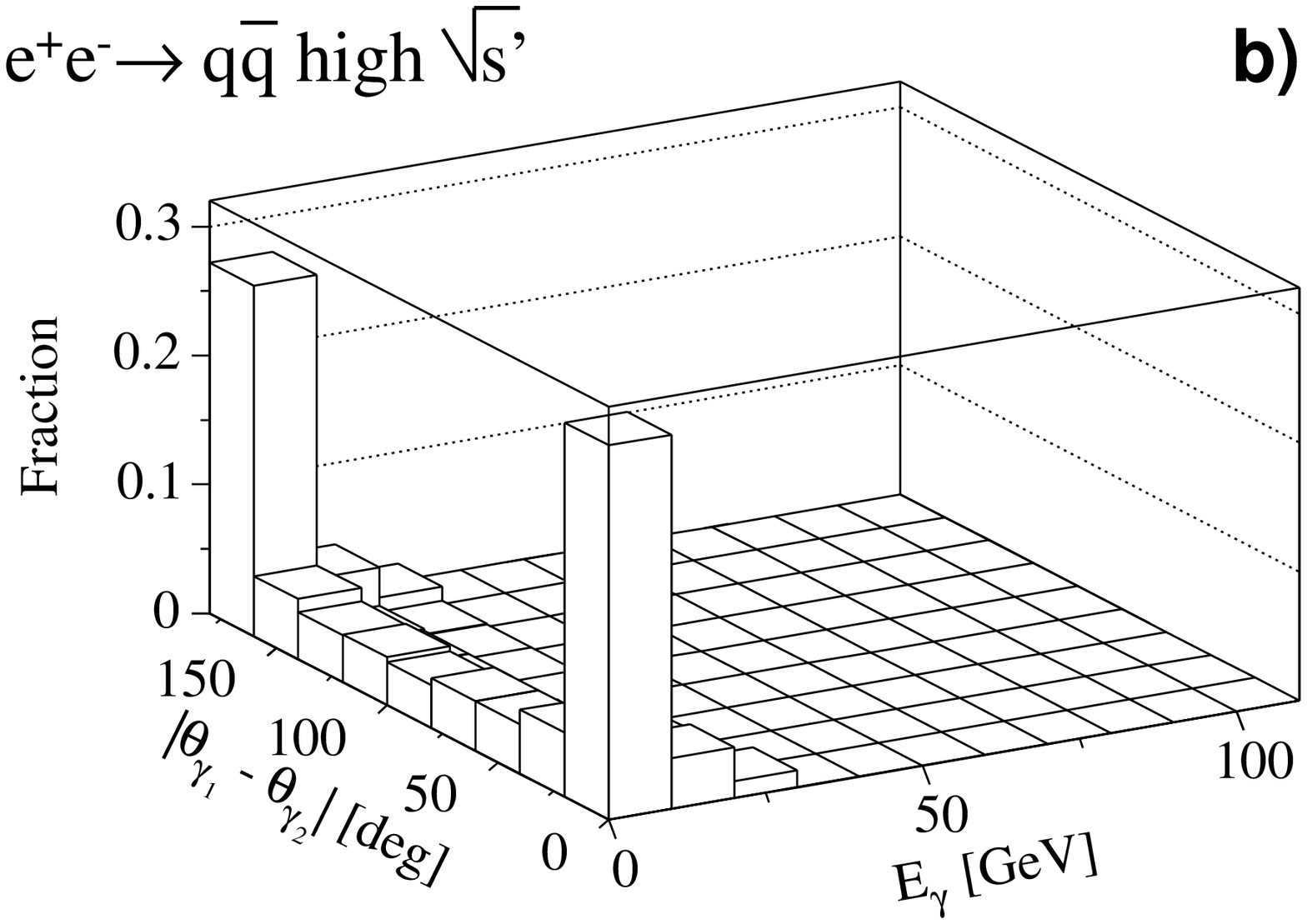}}    \\
       \mbox{\includegraphics[width=.5\textwidth]{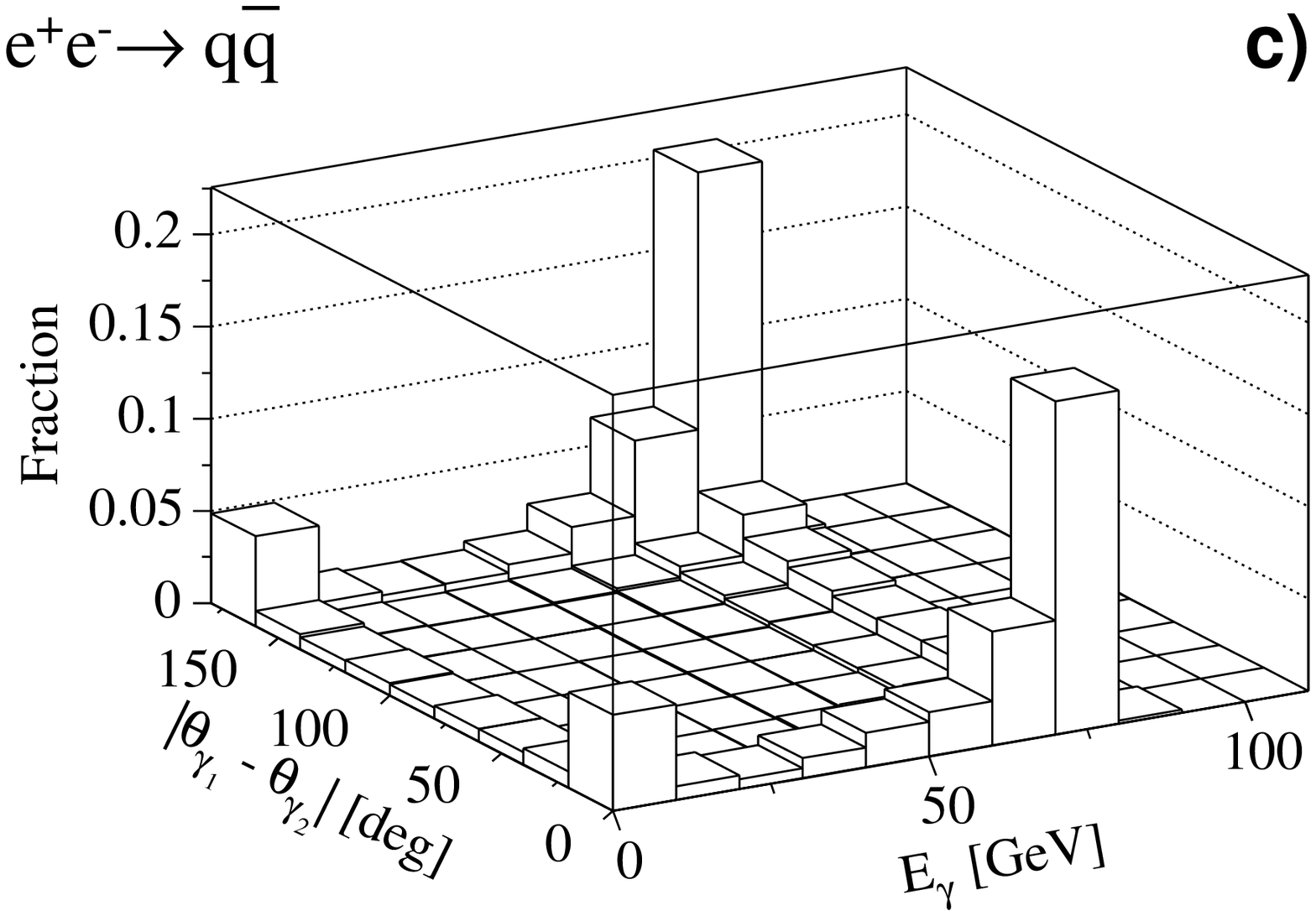}}   &
       \mbox{\includegraphics[width=.5\textwidth]{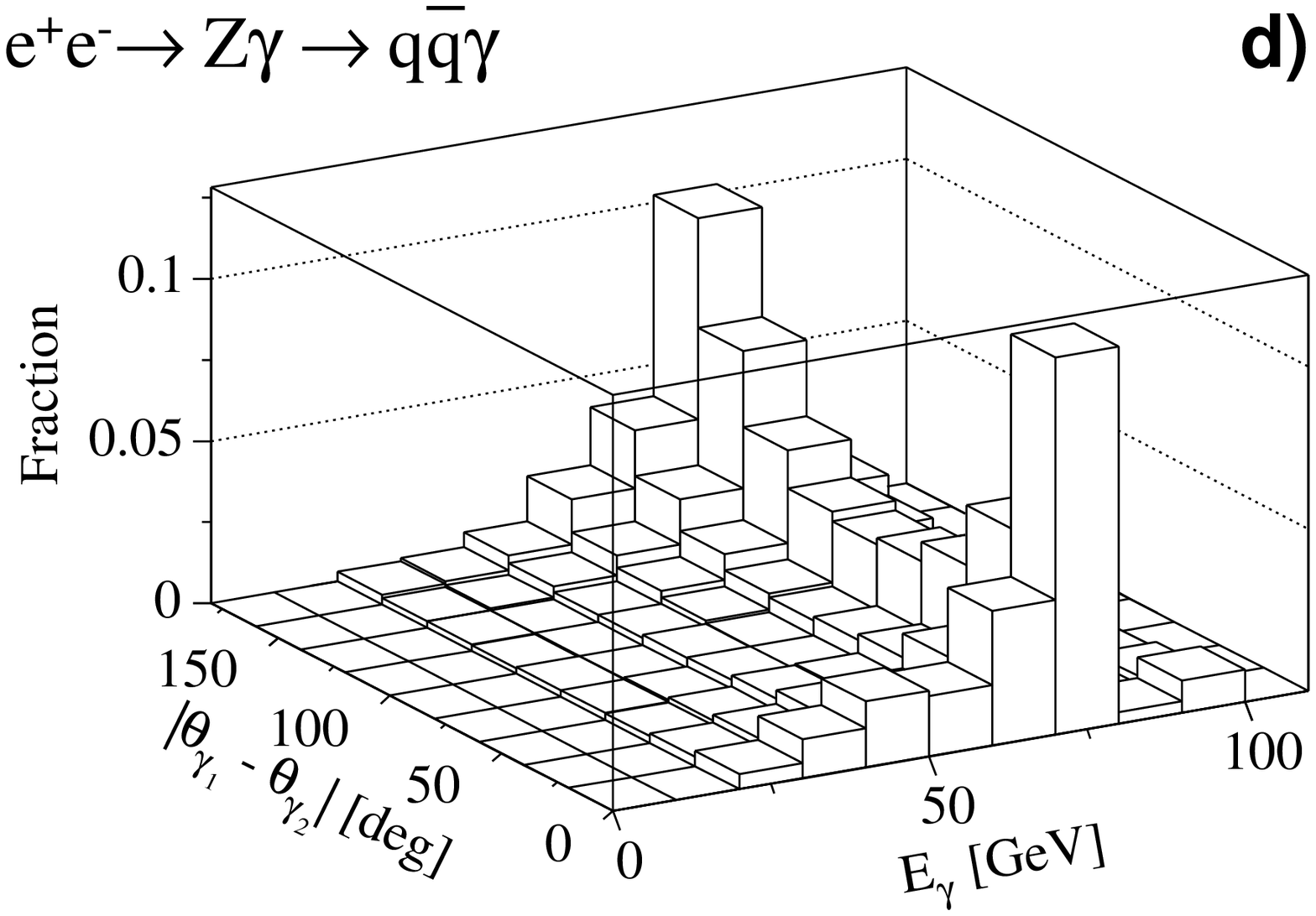}}     \\
       \mbox{\includegraphics[width=.5\textwidth]{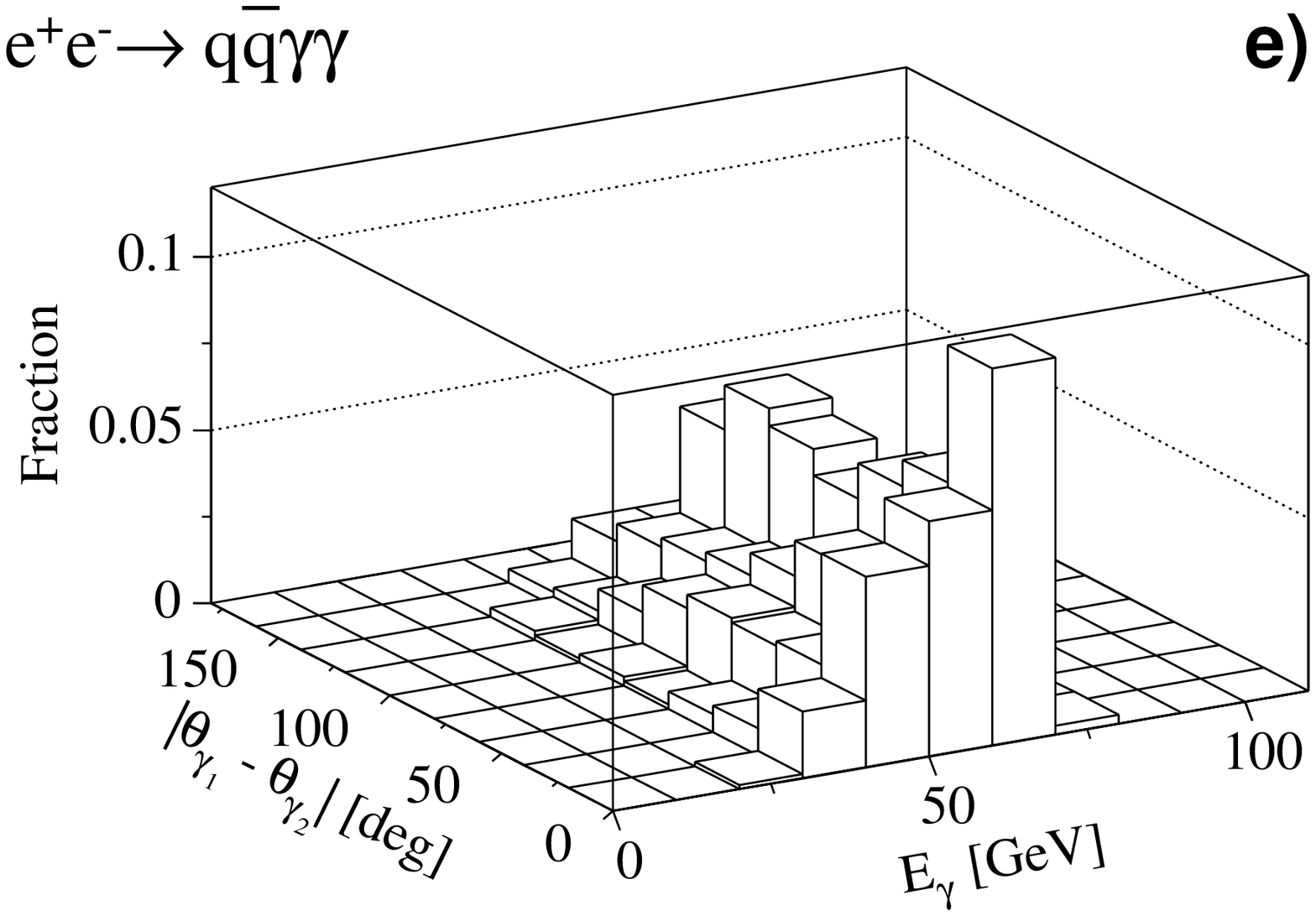}}   & \\
     \end{tabular}
      \icaption{Distributions at generator level of the absolute
	difference of the two photon polar angles versus the energy of
	the most energetic photon for a) the signal, b) the
	$\epem\ra\qqbar$ process for $\sqrt{s'}/\sqrt{s} > 0.85$ and
	$\sqrt{s'}>60\GeV$, c) the $\epem\ra\qqbar$ process for
	$\sqrt{s'}>60\GeV$, d) the $\epem\ra\Zo\gamma\ra\qqbar\gamma$
	process, and e) the
	$\epem\ra\Zo\gamma\gamma\ra\qqbar\gamma\gamma$ process.  Only
	events with at least two photons with energies greater than
	$1\MeV$ are shown, for a sample at $\sqrt{s}=189\GeV$.  The
	histograms show the fraction of the cross section of each
	process in each bin. These cross sections are, respectively,
	5.8~pb, 89.7~pb, 16.1~pb, 20.3~pb and 0.4~pb.
     \label{fig:1}}
  \end{center}
\end{figure}

This analysis selects $\epem\ra\qqbar$ events with two or
more hard ISR photons satisfying the following phase-space criteria:
\begin{equation}
E_{\gamma_{1,2}}> 5 \GeV
\end{equation}
\begin{equation}
|\cos{\theta_{\gamma_{1,2}}}| \ge 0.96
\end{equation}
\begin{equation}
|\sqrt{s'}-m_{\rm Z}|< 2 \Gamma_{\rm Z}
\end{equation}
\begin{equation}
\cos \theta_{\gamma_1} \times \cos \theta_{\gamma_2} < 0\
\end{equation}
\begin{equation}
 \left| p^{\rm T}_{\gamma_1} - p^{\rm T}_{\gamma_2}
  \right|  < 0.1\sqrt{s},
\end{equation}
where $E_{\gamma_i}$, $\theta_{\gamma_i}$ and $p^{\rm T}_{\gamma_i}$
are the energy, polar angle and momentum in the plane transverse to
the beams of the photon $i$, respectively. $\Gamma_{\rm Z}$ denotes
the width of the Z boson, $2.49 \GeV$~\cite{pdg}. If more than two ISR
photons are present in the event, this signal definition is applied to
the two most energetic ones. These criteria select about 6\% of the
phase space of the $\epem\ra\qqbar$ process, corresponding to a cross
section of about 5.5~pb in the $\sqrt{s}$ range explored at LEP.
Figure~\ref{fig:1} illustrates the complementarity of this phase space
with those covered by the analyses of the $\epem\ra\qqbar$,
$\epem\ra\Zo\gamma\ra\qqbar\gamma$ and
$\epem\ra\Zo\gamma\gamma\ra\qqbar\gamma\gamma$ processes described in
References~\citen{l3fpp},~\citen{L3nTGC} and~\citen{L3Zgg}.

%
%%%%%%%%%%%%%%%%%%%%%%%%%%%%%%%%%%%%%%%%%%%%%%%%%%%%%%%%%%%%%%%%%%%%%%%%%%%%%%%
\section{Data and Monte Carlo Samples}
%%%%%%%%%%%%%%%%%%%%%%%%%%%%%%%%%%%%%%%%%%%%%%%%%%%%%%%%%%%%%%%%%%%%%%%%%%%%%%%
%

This measurement is based on 0.62~fb$^{-1}$ of data collected with the
L3 detector~\cite{l3det} at LEP in the years from 1998 through 2000 at
centre-of-mass energies between $\sqrt{s}=188.6\GeV$ and
$\sqrt{s}=209.2\GeV$, as detailed in Table~\ref{tab:data}. In the last
year of data taking, the LEP centre-of-mass energy was routinely
increased while the beams were colliding in order to enhance the
sensitivity of the search for the Standard Model Higgs boson,
exploring the range $\sqrt{s}=202.5-209.2\GeV$. In the following, this
last data sample is split into two energy ranges.

The KK2f~\cite{kk2f} Monte Carlo program is used, with default
options, to generate a total of 1.9 million $\epem\ra\qqbar$ events
which can contain one or more hard ISR photons, at the centre-of-mass
energies listed in Table~\ref{tab:data}. These events correspond to about 35
times the luminosity of the data and cover a phase space much larger
than that of the criteria (1)$-$(5). If at least two ISR photons which
satisfy the criteria (1)$-$(5) are present in an event this is treated
as signal, otherwise it is considered as background. This distinction
between signal and background is performed on generated variables,
before any event simulation and any application of detector
resolutions.

Other background processes are generated with the Monte Carlo programs
PYTHIA~\cite{pythia} for $\rm e^+ e^- \rightarrow Z \epem$ and $\rm
e^+ e^- \rightarrow ZZ$, KK2f for $\rm e^+ e^- \rightarrow \tau^+
\tau^-$, PHOJET~\cite{phojet} for hadron production in two-photon
collisions and KORALW\,~\cite{koralw} for W-boson pair production
except for $\rm e\nu_\e q\bar q'$ final states, generated with
EXCALIBUR~\cite{excalibur}.  The
hadronisation process for signal and background events is modelled
with the JETSET~\cite{pythia} program.

The L3 detector response is simulated using the GEANT~\cite{geant} and
GHEISHA~\cite{gheisha} programs, which model the effects of energy
loss, multiple scattering and showering in the detector.
Time-dependent detector efficiencies, as monitored during
data-taking periods, are also simulated.

%
%%%%%%%%%%%%%%%%%%%%%%%%%%%%%%%%%%%%%%%%%%%%%%%%%%%%%%%%%%%%%%%%%%%%%%%%%%%%%%%
\section{Event Selection}
%%%%%%%%%%%%%%%%%%%%%%%%%%%%%%%%%%%%%%%%%%%%%%%%%%%%%%%%%%%%%%%%%%%%%%%%%%%%%%%
%

\begin{figure}[p]
  \begin{center}
     \begin{tabular}{cc}
       \mbox{\includegraphics[width=.5\textwidth]{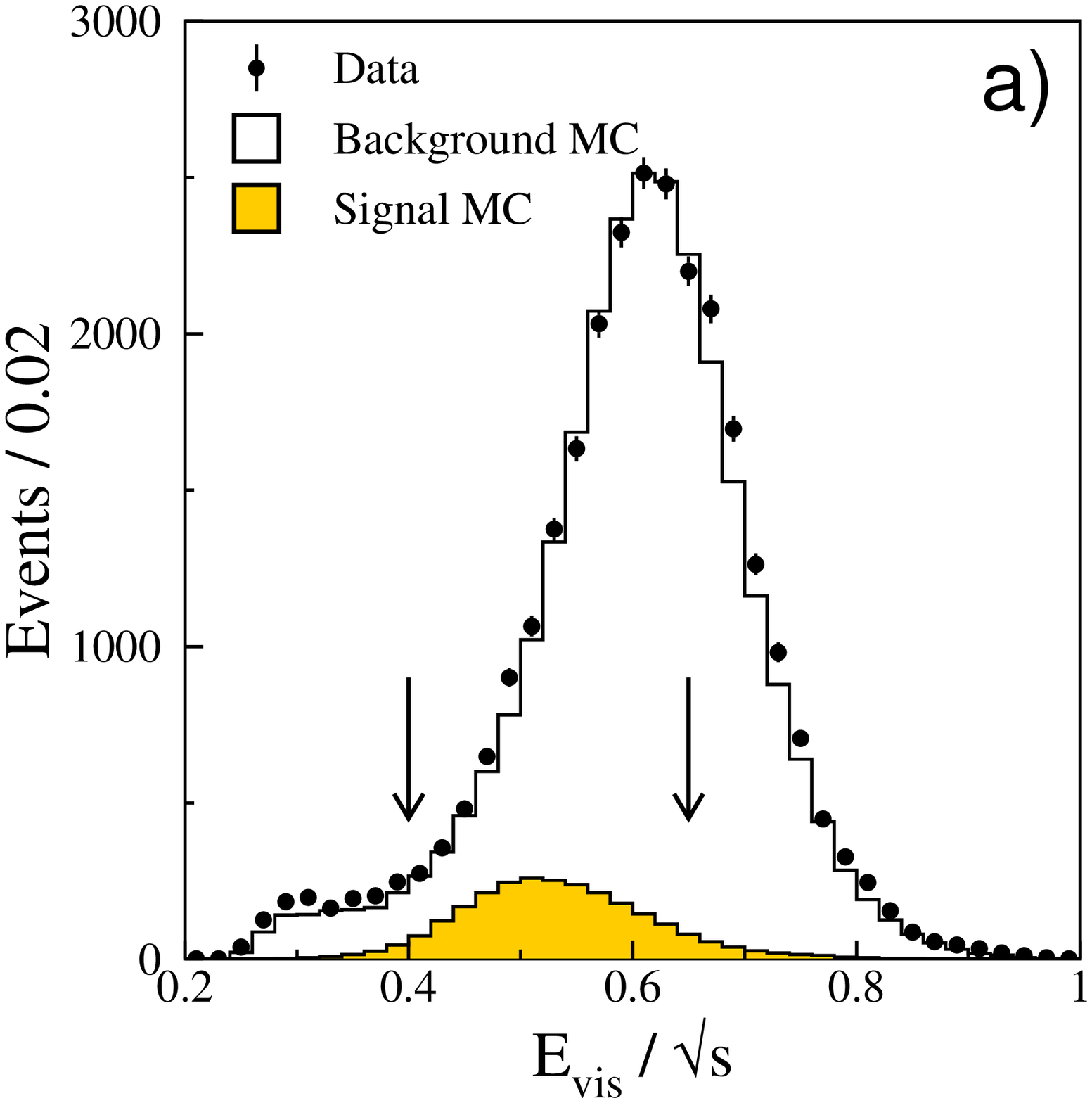}} &
       \mbox{\includegraphics[width=.5\textwidth]{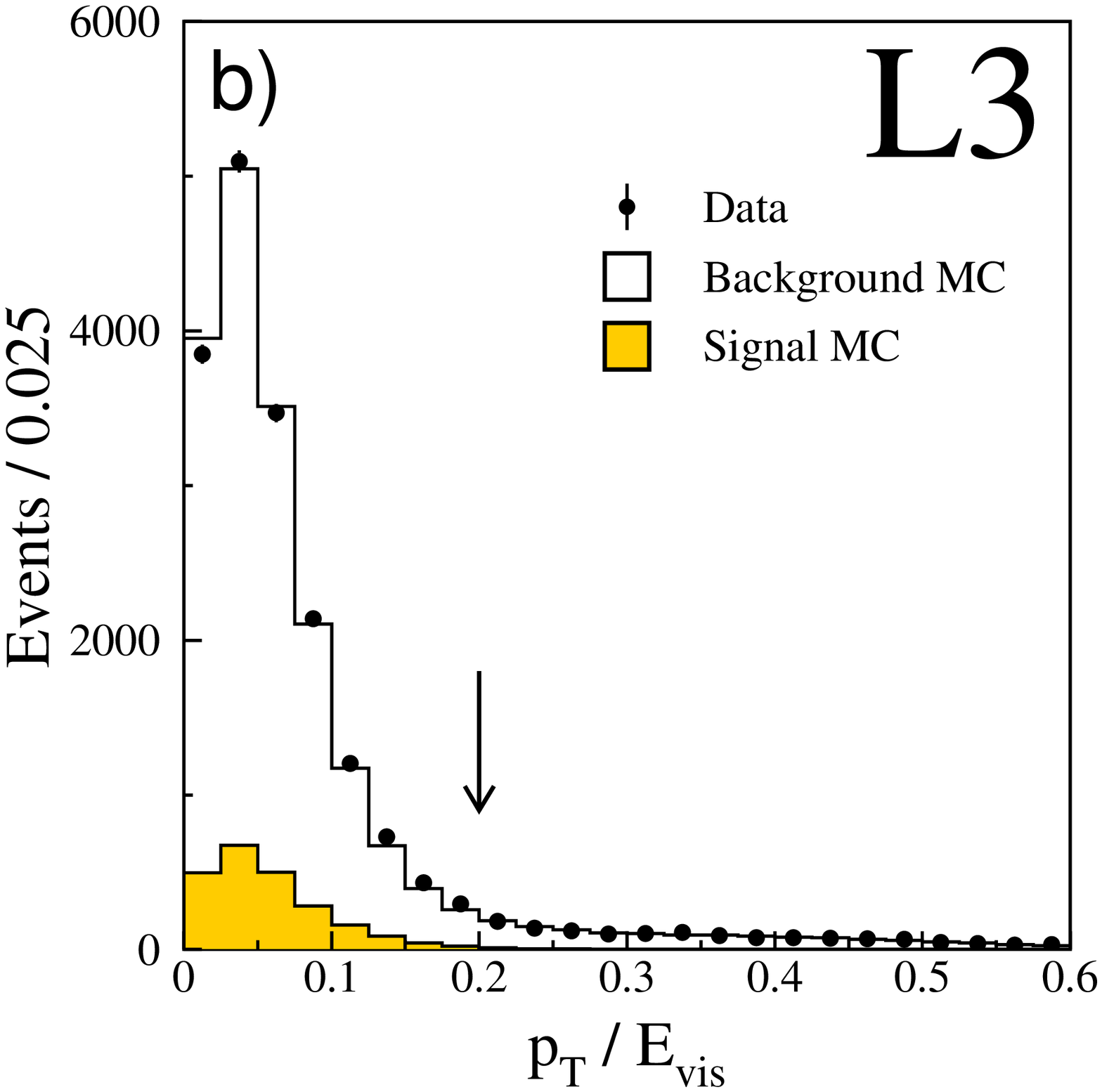}} \\
       \mbox{\includegraphics[width=.5\textwidth]{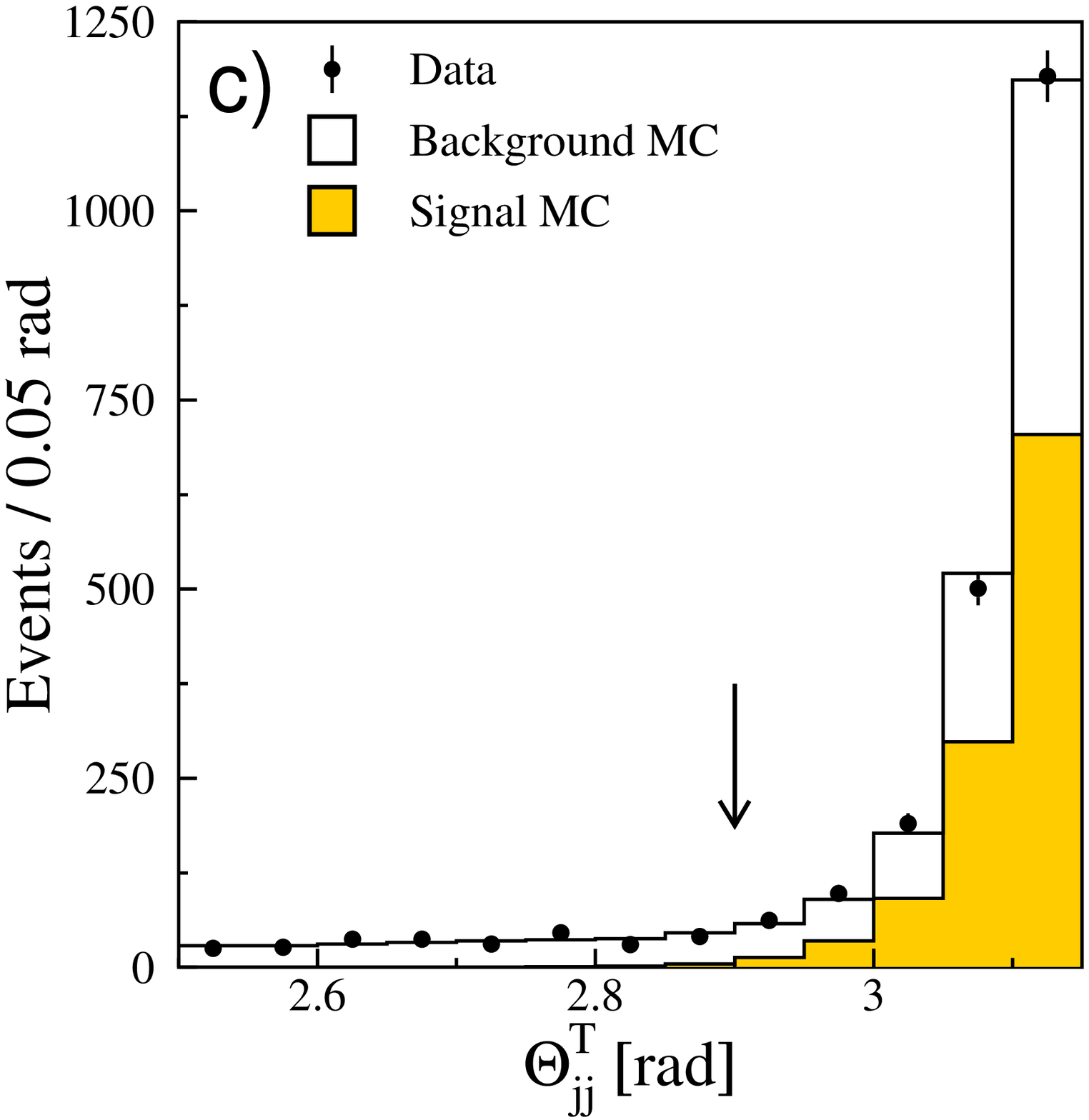}} &
       \mbox{\includegraphics[width=.5\textwidth]{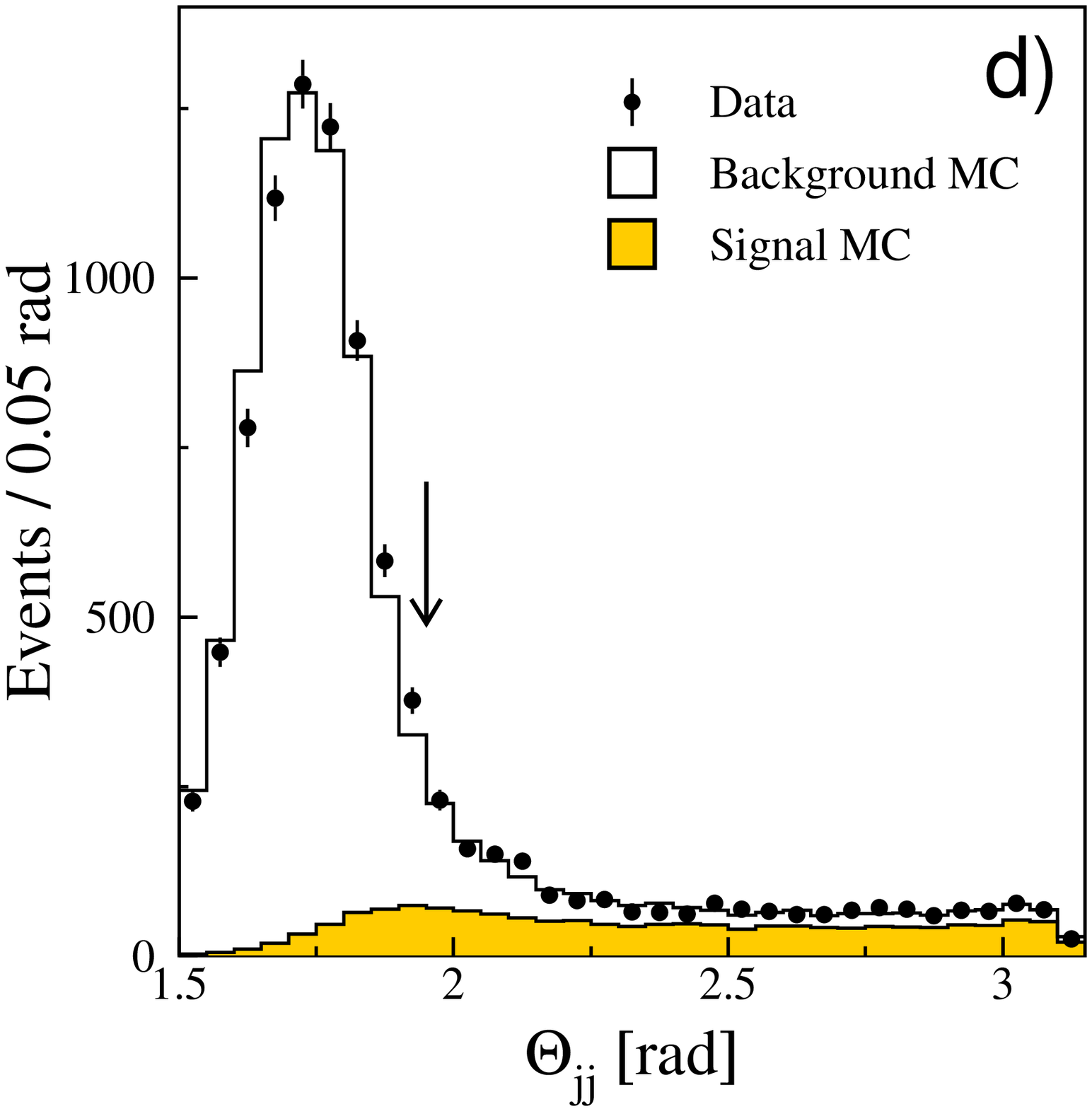}} \\
     \end{tabular}
      \icaption{Distribution for data and Monte Carlo of
       a) the visible energy divided by the centre-of-mass energy;
       b) the sum of the transverse momenta of the two jets  divided by the visible energy;
       c) the angle between the two jets in the plane transverse to the
       beams and 
       d) the angle between the jets. The distributions in a) and b) refer to pre-selection
       level, those in c) and d) to the final selection. The arrows
       represent the position of the cuts, once all other
       pre-selection or selection cuts are applied.
     \label{fig:2}}
  \end{center}
\end{figure}

The event selection proceeds from a sample of high-multiplicity
events.  Events containing photons, electrons or muons with energies
above $20\GeV$ are removed in order to reduce the backgrounds from
$\epem\ra\qqbar$ events with ISR photons in the detector and events
containing W bosons which decay into leptons. The visible mass,
$M_{\rm vis}$, and the visible energy, $E_{\rm vis}$, of these events
are required to satisfy $50\GeV<M_{\rm vis} <140\GeV$ and $0.4<E_{\rm
vis}/\sqrt{s}<0.65$, to reduce both $\epem\ra\qqbar$ events without
missing energy due to ISR photons and most events from two-photon
collisions. The latter cut is illustrated in Figure~\ref{fig:2}a.
Events from two-photon collisions are further suppressed by requiring
$|\cos\theta_{\rm thrust}|<0.96$, where $\theta_{\rm thrust}$ is the
angle between the thrust axis and the beam line. Events are then
reconstructed into two jets by means of the DURHAM
algorithm~\cite{durham} and the signal signature of two back-to-back
jets is enforced by requiring the angle between the two jets,
$\theta_{\rm jj}$, to satisfy $\theta_{\rm jj}>1.5~ {\rm
rad}$. Finally, the sum of the momenta of the two jets in the plane
transverse to the beams, $p_{\rm T}$, must be less than $0.2 E_{\rm
vis}$. This cut, shown in Figure~\ref{fig:2}b, accounts for the fact
that all missing momentum in signal events is due to the two ISR
photons nearly collinear with the beam particles and therefore
directed along the beam line.  After this pre-selection, 17208 events
are selected in data, well consistent with the 17151 events expected
from Monte Carlo simulations, of which 13\% are from signal and 87\%
from background. The background is almost entirely composed by
$\epem\ra\qqbar$ events which do not satisfy the signal definition
(1)$-$(5). Small contributions arise from four-fermion production and
hadron production in two-photon collisions.  The signal efficiency at
this stage of the analysis is 68\%.

\begin{figure}[p]
  \begin{center}
      \mbox{ \includegraphics[width=\textwidth]{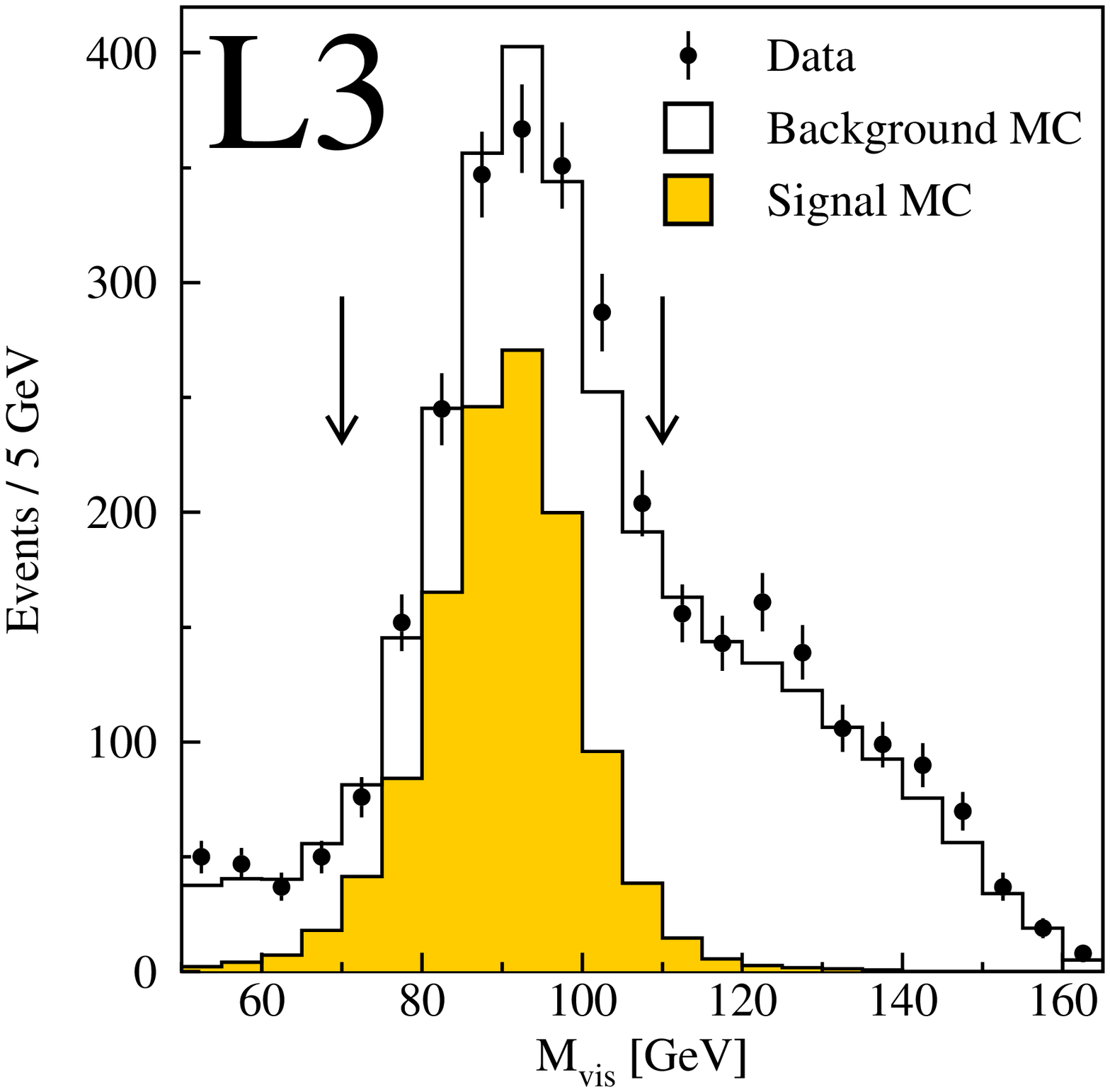}}
      \icaption{Distribution of the visible mass for data and Monte
      Carlo after the application of all other selection cuts.
     \label{fig:3}}
  \end{center}
\end{figure}

Three additional cuts are devised to reduce the residual background
and enhance the signal component in this sample. The energy of the
most energetic jet must be greater than $0.4\sqrt{s}$; the angle
between the two jets in the plane transverse to the beams,
$\theta_{\rm jj}^{\rm T}$, is required to be $\theta_{\rm jj}^{\rm
T}>2.9~{\rm rad}$, as shown in Figure~\ref{fig:2}c; the polar angle of
the jet closest to the beam line, $\theta^{\rm jet}_{\rm low}$, should
be such that $|\cos\theta^{\rm jet}_{\rm low}|<0.85$. Finally, two of
the pre-selection criteria are tightened: $\theta_{\rm jj}>1.95~{\rm
rad}$ and $70\GeV<M_{\rm vis} <110\GeV$, as shown in
Figures~\ref{fig:2}d and~\ref{fig:3}, respectively. The former
criterion is extremely efficient in removing the background from
one-photon radiative return to the Z boson, which is characterised by
a larger boost than the signal and therefore a smaller jet
opening-angle.

After these selection criteria, 1672 events are selected in data while
1684 are expected from Monte Carlo simulations, of which 61\% are from
signal, and 39\% from background, as detailed in
Table~\ref{tab:events}. Three quarters of the background are due to
$\epem\ra\qqbar$ events which do not satisfy the signal definition
(1)$-$(5). The remaining background is due to four-fermion production
and hadron production in two-photon collisions.  The average signal
efficiency is 31\%.

The distribution of $M_{\rm vis}$, shown in Figure~\ref{fig:3}, presents a 
clear enhancement at $m_{\rm Z}$, as expected for signal events.

%
%%%%%%%%%%%%%%%%%%%%%%%%%%%%%%%%%%%%%%%%%%%%%%%%%%%%%%%%%%%%%%%%%%%%%%%%%%%%%%%
\section{Systematic Uncertainties}
%%%%%%%%%%%%%%%%%%%%%%%%%%%%%%%%%%%%%%%%%%%%%%%%%%%%%%%%%%%%%%%%%%%%%%%%%%%%%%%
%

Several sources of possible systematic uncertainties are considered,
and their effects are summarised in Table~\ref{tab:syst}.

Monte Carlo simulations might not perfectly reproduce the tails of the
variables used in the event selection owing to, for instance,
non-linearity in the modelling of the calorimeter response or a bias in
the determination of jet directions close to the edge of fiducial
volumes. To assess this effect, the analysis is repeated by removing
one selection criterion at a time. In addition, a 0.5\% uncertainty in
the jet energy-scale and a 2\% uncertainty in the determination of the
jet angles are also considered.

The signal and background events from the $\epem\ra\qqbar$ process are
generated taking into account the interference between ISR photons and those
emitted in the final state. The analysis is repeated by using a Monte
Carlo sample without this interference and the difference with the
original result is used as an extreme systematic uncertainty on the
modelling of this phenomenon.

The cross sections are measured by assuming a fixed background level,
as discussed below.  Uncertainties in the background cross sections
are therefore a possible source of systematic uncertainty, which is
estimated by repeating the analysis with a variation of 10\% for the
cross section of the $\epem\ra\rm e\nu_\e q\bar q'$ process, 5\% for
$\epem\ra\qqbar$ events classified as background, 5\% for the $\rm e^+
e^- \rightarrow ZZ$ process, 5\% for the $\rm e^+ e^- \rightarrow
Z \epem$ process and 0.5\% for W-boson pair production.

Finally, statistical uncertainties related to the limited amount of
Monte Carlo events used to describe the signal and the background
processes are included as systematic uncertainties. The total
systematic uncertainty on the signal cross section varies between 
5.3\% and 7.7\%, depending on the centre-of-mass energy.

%
%%%%%%%%%%%%%%%%%%%%%%%%%%%%%%%%%%%%%%%%%%%%%%%%%%%%%%%%%%%%%%%%%%%%%%%%%%%%%%%
\section{Results}
%%%%%%%%%%%%%%%%%%%%%%%%%%%%%%%%%%%%%%%%%%%%%%%%%%%%%%%%%%%%%%%%%%%%%%%%%%%%%%%
%

\begin{figure}[p]
  \begin{center}
      \mbox{ \includegraphics[width=\textwidth]{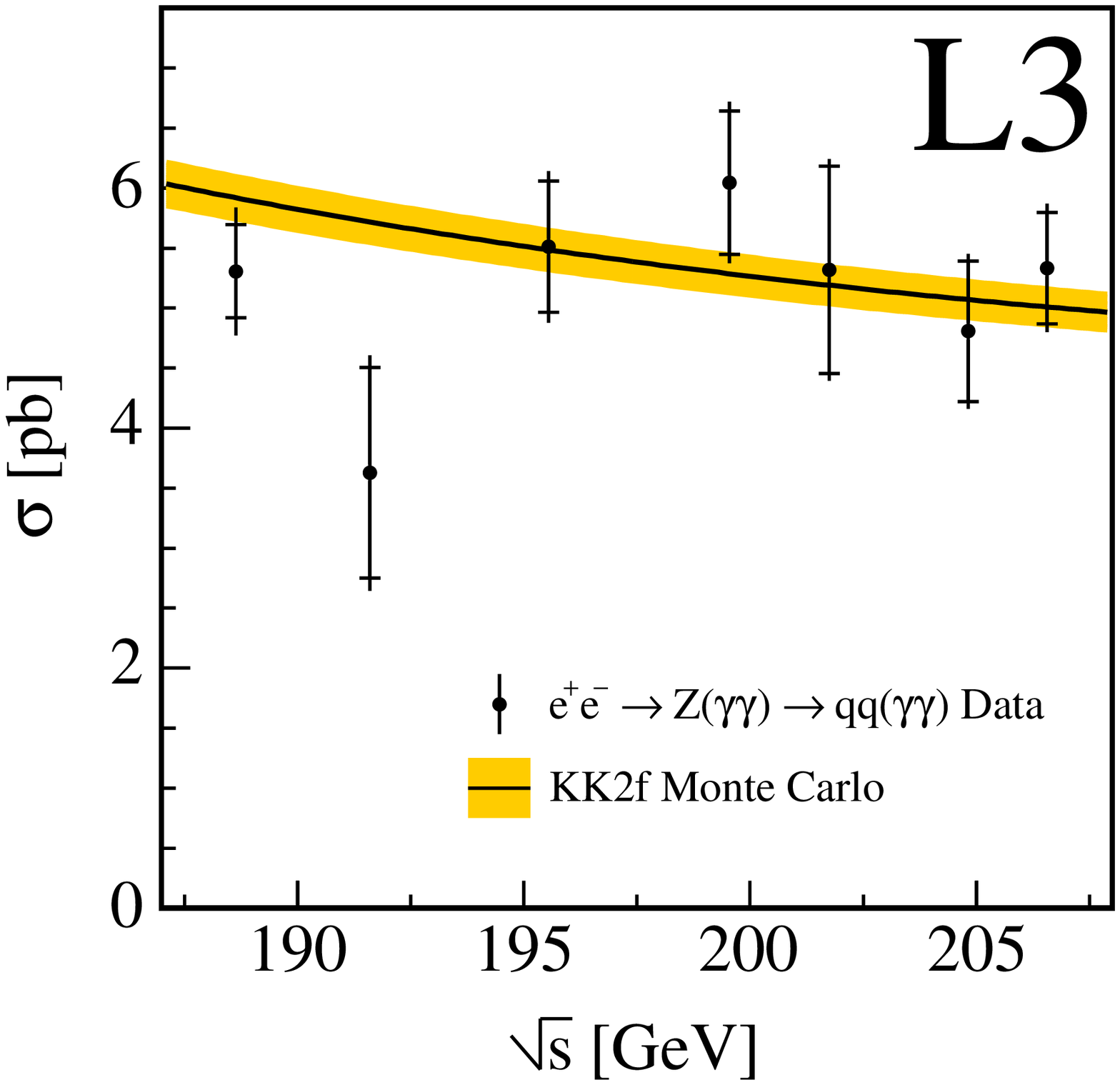}}
      \icaption{Measured cross sections for the various centre-of-mass
      energies, indicated by the points, compared with the Standard
      Model predictions, indicated by the band. The bars on the point
      show the sum in quadrature of the statistical and systematic
      uncertainties. The inner bars represent the statistical
      uncertainties. The width of the band corresponds to an
      uncertainty of 3\% on the predictions, derived as discussed in
      the text.
     \label{fig:4}}
  \end{center}
\end{figure}

The signal cross sections are determined for each centre-of-mass
energy by fitting the observed distributions of $M_{\rm vis}$.
Two components are considered, both with a shape fixed to Monte Carlo
expectations: a signal component with a free normalisation, and a
background component with fixed normalisation. The results are listed
in Table~\ref{tab:events} and plotted in Figure~\ref{fig:4}, together
with the corresponding statistical and systematic uncertainties. A
good agreement with the predictions of the KK2f Monte Carlo, also
given in Table~\ref{tab:events} and Figure~\ref{fig:4}, is
observed. These predictions have an uncertainty of 3\%, which includes
a statistical component and the uncertainty from higher-order
corrections, estimated following the suggestions in
Reference~\citen{kk2f}.

To further compare the results and the expectations, the ratio between
the measured, $\sigma$, and the expected, $\sigma_{\rm th}$, values of
the cross section is calculated for each centre-of-mass energy. These
values are then averaged, by assuming all systematic uncertainties to
be fully correlated, with the exception of those due to the limited
Monte Carlo statistics. The result is:
\begin{displaymath}
  \sigma / \sigma_{\rm th} = 0.98 \pm  0.04 \pm  0.06\,,
\end{displaymath}
where the first uncertainty is statistical and the second
systematic. 

In conclusion, the cross section of the process
$\epem\ra\Zo\gamma\ra\qqbar\gamma\gamma$, where the two photons are
emitted in the phase space defined by the criteria (1)--(5), is measured with an accuracy of
7\% and is well reproduced by the current simulations of ISR in
hadronic events. This finding validates the estimate of the background
from events with two back-to-back jets with mass close to the mass of
the Z boson both in the searches for Higgs bosons of the Standard Model and
beyond and for other manifestations of New Physics.

%
%%%%%%%%%%%%%%%%%%%%%%%%%%%%%%%%%%%%%%%%%%%%%%%%%%%%%%%%%%%%%%%%%%%%%%%%%%%%%%%
%\section*{Author List}
%%%%%%%%%%%%%%%%%%%%%%%%%%%%%%%%%%%%%%%%%%%%%%%%%%%%%%%%%%%%%%%%%%%%%%%%%%%%%%%
%

\newpage
\typeout{   }     
\typeout{Using author list for paper 287 -  }
\typeout{$Modified: Jul 15 2001 by smele $}
\typeout{!!!!  This should only be used with document option a4p!!!!}
\typeout{   }
%
%
%
%  L A T E X  version!!
%
%
% Make sure that the Lep package has been used!
%\input{Lep.sty}%
%
%\ifx\LepCalled\undefined%
%\typeout{     }%
%\typeout{!!!!!!!!!!!!!!!!!!!!!!!!!!!!!!!!!!!!!!!!!!!!!!!!!!!!!!!!!!!}%
%\typeout{Yikes.  You haven't used the Lep package!}%
%\typeout{Please put \protect\usepackage\protect{Lep\protect} in your preamble,
%         followed by}%
%\typeout{\protect\Lep\protect{1\protect} or \protect\Lep\protect{2\protect}}%
%\typeout{     }%
%\typeout{For now you will get a Lep phase 2 authorlist (may not be right!).}%
%\typeout{!!!!!!!!!!!!!!!!!!!!!!!!!!!!!!!!!!!!!!!!!!!!!!!!!!!!!!!!!!!}%
%\typeout{     }%
%\Lep{2}\fi%

\newcount\tutecount  \tutecount=0
\def\tutenum#1{\global\advance\tutecount by 1 \xdef#1{\the\tutecount}}
\def\tute#1{$^{#1}$}
\tutenum\aachen            % 1 
\tutenum\nikhef            % 2 
\tutenum\mich              % 3 
\tutenum\lapp              % 4 
\tutenum\basel             % 5 
\tutenum\lsu               % 6 
\tutenum\beijing           % 7 
\tutenum\bologna           % 8 
\tutenum\tata              % 9 
\tutenum\ne                % 10
\tutenum\bucharest         % 11
\tutenum\budapest          % 12
\tutenum\mit               % 13
\tutenum\panjab            % 14 
\tutenum\debrecen          % 15
\tutenum\dublin            % 16
\tutenum\florence          % 17
\tutenum\cern              % 18
\tutenum\wl                % 19
\tutenum\geneva            % 20
\tutenum\hamburg           % 21
\tutenum\hefei             % 22
\tutenum\lausanne          % 23
\tutenum\lyon              % 24
\tutenum\madrid            % 25
\tutenum\florida           % 26
\tutenum\milan             % 27
\tutenum\moscow            % 29
\tutenum\naples            % 30
\tutenum\cyprus            % 31
\tutenum\nymegen           % 32
\tutenum\caltech           % 33
\tutenum\perugia           % 34
\tutenum\peters            % 35
\tutenum\cmu               % 36
\tutenum\potenza           % 37
\tutenum\prince            % 38
\tutenum\riverside         % 39
\tutenum\rome              % 40
\tutenum\salerno           % 41
\tutenum\ucsd              % 42
\tutenum\sofia             % 43
\tutenum\korea             % 44
\tutenum\taiwan            % 45
\tutenum\tsinghua          % 46
\tutenum\purdue            % 47
\tutenum\psinst            % 48
\tutenum\zeuthen           % 49
\tutenum\eth               % 50

{
\parskip=0pt
\noindent
{\bf The L3 Collaboration:}
\ifx\selectfont\undefined%  old style font selection
 \baselineskip=10.8pt
 \baselineskip\baselinestretch\baselineskip
 \normalbaselineskip\baselineskip
 \ixpt
\else%                      new style font selection
 \fontsize{9}{10.8pt}\selectfont
\fi
\medskip
\tolerance=10000
\hbadness=5000
\raggedright
\hsize=162truemm\hoffset=0mm
\def\r{\rlap,}
\noindent

P.Achard\r\tute\geneva\ 
O.Adriani\r\tute{\florence}\ 
M.Aguilar-Benitez\r\tute\madrid\ 
J.Alcaraz\r\tute{\madrid}\ 
G.Alemanni\r\tute\lausanne\
J.Allaby\r\tute\cern\
A.Aloisio\r\tute\naples\ 
M.G.Alviggi\r\tute\naples\
H.Anderhub\r\tute\eth\ 
V.P.Andreev\r\tute{\lsu,\peters}\
% ...
\newpage
%\rule{\textwidth}{0.4pt}
\begin{list}{A}{\itemsep=0pt plus 0pt minus 0pt\parsep=0pt plus 0pt minus 0pt
                \topsep=0pt plus 0pt minus 0pt}
\item[\aachen]
 III. Physikalisches Institut, RWTH, D-52056 Aachen, Germany$^{\S}$
\item[\nikhef] National Institute for High Energy Physics, NIKHEF, 
     and University of Amsterdam, NL-1009 DB Amsterdam, The Netherlands
\item[\mich] University of Michigan, Ann Arbor, MI 48109, USA
\item[\lapp] Laboratoire d'Annecy-le-Vieux de Physique des Particules, 
     LAPP,IN2P3-CNRS, BP 110, F-74941 Annecy-le-Vieux CEDEX, France
\item[\basel] Institute of Physics, University of Basel, CH-4056 Basel,
     Switzerland
\item[\lsu] Louisiana State University, Baton Rouge, LA 70803, USA
\item[\beijing] Institute of High Energy Physics, IHEP, 
  100039 Beijing, China$^{\triangle}$ 
\item[\bologna] University of Bologna and INFN-Sezione di Bologna, 
     I-40126 Bologna, Italy
\item[\tata] Tata Institute of Fundamental Research, Mumbai (Bombay) 400 005, India
\item[\ne] Northeastern University, Boston, MA 02115, USA
\item[\bucharest] Institute of Atomic Physics and University of Bucharest,
     R-76900 Bucharest, Romania
\item[\budapest] Central Research Institute for Physics of the 
     Hungarian Academy of Sciences, H-1525 Budapest 114, Hungary$^{\ddag}$
\item[\mit] Massachusetts Institute of Technology, Cambridge, MA 02139, USA
\item[\panjab] Panjab University, Chandigarh 160 014, India
\item[\debrecen] KLTE-ATOMKI, H-4010 Debrecen, Hungary$^\P$
\item[\dublin] Department of Experimental Physics,
  University College Dublin, Belfield, Dublin 4, Ireland
\item[\florence] INFN Sezione di Firenze and University of Florence, 
     I-50125 Florence, Italy
\item[\cern] European Laboratory for Particle Physics, CERN, 
     CH-1211 Geneva 23, Switzerland
\item[\wl] World Laboratory, FBLJA  Project, CH-1211 Geneva 23, Switzerland
\item[\geneva] University of Geneva, CH-1211 Geneva 4, Switzerland
\item[\hamburg] University of Hamburg, D-22761 Hamburg, Germany
\item[\hefei] Chinese University of Science and Technology, USTC,
      Hefei, Anhui 230 029, China$^{\triangle}$
\item[\lausanne] University of Lausanne, CH-1015 Lausanne, Switzerland
\item[\lyon] Institut de Physique Nucl\'eaire de Lyon, 
     IN2P3-CNRS,Universit\'e Claude Bernard, 
     F-69622 Villeurbanne, France
\item[\madrid] Centro de Investigaciones Energ{\'e}ticas, 
     Medioambientales y Tecnol\'ogicas, CIEMAT, E-28040 Madrid,
     Spain${\flat}$ 
\item[\florida] Florida Institute of Technology, Melbourne, FL 32901, USA
\item[\milan] INFN-Sezione di Milano, I-20133 Milan, Italy
\item[\moscow] Institute of Theoretical and Experimental Physics, ITEP, 
     Moscow, Russia
\item[\naples] INFN-Sezione di Napoli and University of Naples, 
     I-80125 Naples, Italy
\item[\cyprus] Department of Physics, University of Cyprus,
     Nicosia, Cyprus
\item[\nymegen] Radboud University and NIKHEF, 
     NL-6525 ED Nijmegen, The Netherlands
\item[\caltech] California Institute of Technology, Pasadena, CA 91125, USA
\item[\perugia] INFN-Sezione di Perugia and Universit\`a Degli 
     Studi di Perugia, I-06100 Perugia, Italy   
\item[\peters] Nuclear Physics Institute, St. Petersburg, Russia
\item[\cmu] Carnegie Mellon University, Pittsburgh, PA 15213, USA
\item[\potenza] INFN-Sezione di Napoli and University of Potenza, 
     I-85100 Potenza, Italy
\item[\prince] Princeton University, Princeton, NJ 08544, USA
\item[\riverside] University of Californa, Riverside, CA 92521, USA
\item[\rome] INFN-Sezione di Roma and University of Rome, ``La Sapienza",
     I-00185 Rome, Italy
\item[\salerno] University and INFN, Salerno, I-84100 Salerno, Italy
\item[\ucsd] University of California, San Diego, CA 92093, USA
\item[\sofia] Bulgarian Academy of Sciences, Central Lab.~of 
     Mechatronics and Instrumentation, BU-1113 Sofia, Bulgaria
\item[\korea]  The Center for High Energy Physics, 
     Kyungpook National University, 702-701 Taegu, Republic of Korea
\item[\taiwan] National Central University, Chung-Li, Taiwan, China
\item[\tsinghua] Department of Physics, National Tsing Hua University,
      Taiwan, China
\item[\purdue] Purdue University, West Lafayette, IN 47907, USA
\item[\psinst] Paul Scherrer Institut, PSI, CH-5232 Villigen, Switzerland
\item[\zeuthen] DESY, D-15738 Zeuthen, Germany
\item[\eth] Eidgen\"ossische Technische Hochschule, ETH Z\"urich,
     CH-8093 Z\"urich, Switzerland
\item[\S]  Supported by the German Bundesministerium 
        f\"ur Bildung, Wissenschaft, Forschung und Technologie.
\item[\ddag] Supported by the Hungarian OTKA fund under contract
numbers T019181, F023259 and T037350.
\item[\P] Also supported by the Hungarian OTKA fund under contract
  number T026178.
\item[$\flat$] Supported also by the Comisi\'on Interministerial de Ciencia y 
        Tecnolog{\'\i}a.
\item[$\sharp$] Also supported by CONICET and Universidad Nacional de La Plata,
        CC 67, 1900 La Plata, Argentina.
\item[$\triangle$] Supported by the National Natural Science
  Foundation of China.
\end{list}
}
\vfill

%%% Local Variables: 
%%% mode: latex
%%% TeX-master: t
%%% End:

\newpage

%
%%%%%%%%%%%%%%%%%%%%%%%%%%%%%%%%%%%%%%%%%%%%%%%%%%%%%%%%%%%%%%%%%%%%%%%%%%%%%%%

\newpage

%
%%%%%%%%%%%%%%%%%%%%%%%%%%%%%%%%%%%%%%%%%%%%%%%%%%%%%%%%%%%%%%%%%%%%%%%%%%%%%%%
% Tables
%%%%%%%%%%%%%%%%%%%%%%%%%%%%%%%%%%%%%%%%%%%%%%%%%%%%%%%%%%%%%%%%%%%%%%%%%%%%%%%
%

\begin{table}[h]
\begin{center}
\begin{tabular}{|c|c|c|c|c|c|c|c|}
      \hline
$\sqrt{s}$ (Ge\kern -0.1em V) &   188.6 &  191.6 & 195.6 &  199.5 &  201.7 & $202.5-205.5$ &  $205.5-209.2$ \\   
\hline
$\cal{L}$ (pb$^{-1})$ & 176.0 &  29.5 & 83.4 &  81.4 &  36.7 &  77.5 &  138.6  \\ 
      \hline
\end{tabular}
\end{center}
        \icaption[]{\label{tab:data}
        Centre-of-mass energies and 
        corresponding integrated luminosities, $\cal{L}$, considered
        in this analysis. The last two energy ranges correspond to
        the average centre-of-mass energy values $<\sqrt{s}>=204.8 \GeV$
        and $<\sqrt{s}>=206.6 \GeV$, respectively. 
} 
\end{table}

%----------------------------------------------------------------------

\begin{table}[h]
\begin{center}
\begin{tabular}{|c|c|c|c|c|c|c|c|}
\hline
$\sqrt{s}$   &  $N_{\rm Data}$  &  $N_{\rm MC}$          &  $N_{\rm Sign}$            &  $N_{\rm Back}$        &  $\varepsilon$  &  $\sigma$       & $\sigma_{\rm th}$\\
(Ge\kern -0.1em V)     &            &                  &                   &               &   (\%)          &  (pb)           & (pb)         \\\hline
188.6              &  632   &  $ 660.1 \pm 6.9$   &  $372.3 \pm 4.6$   &  $287.9 \pm 5.1$   &  $36.1 \pm 0.4$         &  $5.31 \pm 0.39 \pm 0.37$ &   5.9 \\        
191.6              &  \phantom{0}84    &  $ \phantom{0}97.3 \pm 2.1$   &  $ \phantom{0}56.7 \pm 1.5$   &  $ \phantom{0}40.6 \pm 1.5$   &  $ 34.7 \pm 0.9$         &  $3.63 \pm 0.88 \pm 0.44$ &   5.7 \\        
195.5              &  238   &  $ 236.5 \pm 4.2$  &  $ 149.7 \pm 2.8$  &  $ \phantom{0}86.8 \pm 3.1$   &  $ 32.3 \pm 0.6$         &  $5.51 \pm 0.55 \pm 0.32$ &   5.5 \\        
199.5              &  216   &  $ 196.3 \pm 3.9$  &  $ 126.6 \pm 2.5$  &  $ \phantom{0}69.6 \pm 3.0$   &  $ 29.6 \pm 0.6$         &  $6.05 \pm 0.60 \pm 0.31$ &   5.3 \\        
201.7              &  \phantom{0}82    &  $ \phantom{0}80.9 \pm 2.1$   &  $ \phantom{0}52.8 \pm 1.6$   &  $ \phantom{0}28.1 \pm 1.4$   &  $ 27.8 \pm 0.8$         &  $5.32 \pm 0.87 \pm 0.33$ &   5.2 \\        
204.8              &  147   &  $ 153.2 \pm 1.6$  &  $ 102.2 \pm 1.0$  &  $ \phantom{0}50.9 \pm 1.2$   &  $ 26.4 \pm 0.3$         &  $4.81 \pm 0.58 \pm 0.27$ &   5.1 \\        
206.6              &  273   &  $ 259.9 \pm 2.3$  &  $ 173.5 \pm 1.3$  &  $ \phantom{0}86.5 \pm 2.0 $  &  $ 25.5 \pm 0.2$         &  $5.33 \pm 0.46 \pm 0.27$ &   5.0 \\ \hline 
\end{tabular}
\caption[]{Number of events observed in data after the final
  selection, $N_{\rm Data}$, compared with the total number of events
  expected from Monte Carlo, $N_{\rm MC}$. The number of signal events
  expected from the KK2f Monte Carlo, $N_{\rm Sign}$, is also given,
  together with the number of background events, $N_{\rm Back}$.  The
  selection efficiency, $\varepsilon$, is also listed, together with
  the measured, $\sigma$, and expected, $\sigma_{\rm th}$, signal
  cross sections. The uncertainties on $N_{\rm MC}$, $N_{\rm Sign}$,
  $N_{\rm Back}$ and $\varepsilon$ correspond to the statistical
  uncertainty of the Monte Carlo. The first uncertainty on $\sigma$ is
  statistical, the second systematic.\\ }
\label{tab:events}
\end{center}
\end{table}

\begin{table}[h]
\begin{center}
\begin{tabular}{|l|c|}
\hline
Source                              & Effect (\%)   \\ \hline 
Selection criteria                  & 2.3 \\ 
Jet energy scale                    & 3.0 \\ 
Jet angle                           & 3.0 \\ 
ISR/FSR interference                & 1.2 \\ 
Background normalisation            & $1.4 - 4.4$ \\
Monte Carlo statistics              & $1.2 - 3.9$ \\ \hline
Total                               & $5.3 - 7.7$ \\ \hline
\end{tabular}
\caption[]{Systematic uncertainties on the signal cross section.
}
\label{tab:syst}
\end{center}
\end{table}

\clearpage

%
%%%%%%%%%%%%%%%%%%%%%%%%%%%%%%%%%%%%%%%%%%%%%%%%%%%%%%%%%%%%%%%%%%%%%%%%%%%%%%%
% Figures
%%%%%%%%%%%%%%%%%%%%%%%%%%%%%%%%%%%%%%%%%%%%%%%%%%%%%%%%%%%%%%%%%%%%%%%%%%%%%%%
%

\newpage
               
\begin{thebibliography}{99} 
%%%%%%%%%%%%%%%%%%%%%%%%%%%%%%%%%%%%%%%%%%%%%%%%%%%%%%%%%%%%%%%%%%%%%%%%%%%%%%
%

\bibitem{l3fpp}
L3 Collab., M. Acciarri \etal, Phys. Lett. {\bf B 479} (2000) 101; \\
L3 Collab., P.~Achard \etal, {\it Measurement of Hadron and Lepton Pair Production at $192\GeV<\sqrt{s}<209\GeV$ at LEP}, in preparation.

\bibitem{othersFpp}
ALEPH Collab.,  R.~Barate \etal,  Eur. Phys. J. {\bf C 12} (2000) 183;\\
DELPHI Collab., P.~Abreu \etal, Phys. Lett. {\bf B 485} (2000) 45;\\
OPAL Collab., G.~Abbiendi \etal, Eur. Phys. J. {\bf C 33} (2004) 173. 

\bibitem{pdg}
S.~Eidelman \etal, Phys. Lett. {\bf B 592} (2004) 1.

\bibitem{L3nTGC}
L3 Collab., P.~Achard \etal, Phys. Lett. {\bf B 597}  (2004) 119.

\bibitem{mZ}
L3 Collab., P.~Achard \etal, Phys. Lett. {\bf B 585}  (2004) 42;\\
OPAL Collab., G.~Abbiendi \etal, Phys. Lett. {\bf B 604}  (2004) 31. 

\bibitem{L3ZggFirst}
L3 Collab., M.~Acciarri \etal, Phys. Lett. {\bf B 478}  (2000) 39.

\bibitem{L3Zgg}
L3 Collab., M.~Acciarri \etal, Phys. Lett. {\bf B 505}  (2001) 47;\\
L3 Collab., P.~Achard \etal, Phys. Lett. {\bf B 540}  (2002) 43;\\
OPAL Collab., G.~Abbiendi \etal, Phys. Rev. {\bf D 70} (2004) 032005.

\bibitem{l3det} 
L3 Collab., B. Adeva {\it et al.}, Nucl. Instr. Meth. {\bf A 289} (1990) 35; \\
L3 Collab., O. Adriani \etal, Phys. Rep. {\bf 236} (1993) 1; \\
J.A. Bakken {\it et al.}, Nucl. Instr. Meth. {\bf A 275} (1989) 81; \\
O. Adriani {\it et al.}, Nucl. Instr. Meth. {\bf A 302} (1991) 53; \\
B. Adeva {\it et al.}, Nucl. Instr. Meth. {\bf A 323} (1992) 109; \\
K. Deiters {\it et al.}, Nucl. Instr. Meth. {\bf A 323} (1992) 162; \\
M. Chemarin {\it et al.}, Nucl. Instr. Meth. {\bf A 349} (1994) 345; \\
M. Acciarri {\it et al.}, Nucl. Instr. Meth. {\bf A 351} (1994) 300; \\
G. Basti {\it et al.}, Nucl. Instr. Meth. {\bf A 374} (1996) 293; \\
A. Adam {\it et al.}, Nucl. Instr. Meth. {\bf A 383} (1996) 342.

\bibitem{kk2f}
KK2f version 4.13 is used; \\
S.~Jadach, B.F.L.~Ward and Z.~W\c{a}s, Comp. Phys. Comm. {\bf 130} (2000) 260.

\bibitem{pythia} 
JETSET version 7.4 and PYTHIA versions 5.722 and 6.1 are used; \\
T. Sj{\"o}strand, preprint CERN-TH/7112/93 (1993), revised 1995; \\
T. Sj{\"o}strand, {Comp. Phys. Comm.} {\bf 82} (1994) 74; \\
T. Sj{\"o}strand, {Comp. Phys. Comm.} {\bf 135} (2001) 238; \\
T. Sj{\"o}strand, preprint hep-ph/0108264 (2001).

\bibitem{phojet}
PHOJET version 1.05 is used; \\
R.~Engel, Z. Phys. {\bf C 66} (1995) 203; \\
R.~Engel and J.~Ranft, {Phys. Rev.} {\bf D 54} (1996) 4244.

\bibitem{koralw} 
KORALW version 1.33 is used; \\
M. Skrzypek {\it et al.}, Comp. Phys. Comm. {\bf 94} (1996) 216;\\
M. Skrzypek {\it et al.}, Phys. Lett. {\bf B 372} (1996) 289.

\bibitem{excalibur}
EXCALIBUR version 1.11 is used; \\
F.A.~Berends, R. Kleiss and R. Pittau, Comp. Phys. Comm. {\bf 85} (1995) 437.

\bibitem{geant}
GEANT version 3.15 is used; \\
R. Brun \etal, preprint CERN DD/EE/84-1 (1985), revised 1987. 

\bibitem{gheisha} H. Fesefeldt, RWTH Aachen Report PITHA 85/02 (1985).

\bibitem{durham} 
S. Catani {\it et al.}, Phys. Lett. {\bf  B 269} (1991) 432;\\
S. Bethke {\it et al.}, Nucl. Phys. {\bf B 370} (1992) 310.


\end{thebibliography}
\end{document}